\documentclass[12pt,preprint]{aastex}
\def\MCN{\mbox{CH$_3$CN}}

\def\HII{H{\sc ii}}

\def\kms{\mbox{km~s$^{-1}$}}

\def\Vlsr{$V_{\rm LSR}$}

\def\Mgas{\mbox{$M_{\rm gas}$}}
\def\Mdyn{\mbox{$M_{\rm dyn}$}}
\def\tff{\mbox{$t_{\rm ff}$}}
\def\tacc{\mbox{$t_{\rm acc}$}}
\def\tout{\mbox{$t_{\rm out}$}}
\def\Macc{\mbox{$\dot{M}_{\rm acc}$}}
\def\Mout{\mbox{$\dot{M}_{\rm out}$}}
\def\vrot{\mbox{$v_{\rm rot}$}}
\def\vin{\mbox{$v_{\rm in}$}}

\received{2003 October 29}
\begin{document}
\title{Rotating disks in high-mass young stellar objects$^\star$}
\slugcomment{$^\star$Based on observations carried out with the IRAM Plateau de Bure
Interferometer. IRAM is supported by INSU/CNRS (France), MPG (Germany) and
IGN (Spain).}

\author{M.\ T.\ Beltr\'an\altaffilmark{1}, R.\ Cesaroni\altaffilmark{1}, R.\ Neri\altaffilmark{2},
C.\ Codella\altaffilmark{3}, R.\ S.\ Furuya\altaffilmark{1,4}, \\
L.\ Testi\altaffilmark{1}, and L. Olmi\altaffilmark{3}}

\altaffiltext{1}{Osservatorio Astrofisico di Arcetri, INAF, Largo E. Fermi 5,
           I-50125 Firenze, Italy;
	   \tt{mbeltran@arcetri.astro.it}, \tt{cesa@arcetri.astro.it}, \tt{lt@arcetri.astro.it}}
\altaffiltext{2}{IRAM, 300 Rue de la Piscine, F-38406 Saint Martin
	   d'H\`eres, France; \tt{neri@iram.fr}} 
\altaffiltext{3}{Istituto di Radioastronomia, CNR, Sezione di Firenze, Largo E. Fermi 5, I-50125 Firenze, Italy; \tt{codella@arcetri.astro.it}, \tt{olmi@arcetri.astro.it}}
\altaffiltext{4}{Division of Physics, Mathematics, and Astronomy, California Institute of Technology, MS 105-24, Pasadena, CA 91125; \tt{rsf@astro.caltech.edu}}
 
\shorttitle{Disks in high-mass young stellar objects}
\shortauthors{Beltr\'an et al.}

\begin{abstract}
We report on the detection of four rotating massive disks in two regions of
high-mass star formation. The disks are perpendicular to known bipolar
outflows and turn out to be unstable but long lived. We infer that
accretion onto the embedded (proto)stars must proceed through the disks with
rates of $\sim$$10^{-2}~M_\odot$~yr$^{-1}$.
\end{abstract}


\section{Introduction}

The formation of massive stars represents a puzzle from a theoretical point
of view. Unlike their low-mass counterparts, they are believed to reach the
zero-age main sequence still deeply embedded in their parental cores:
in particular, Palla \& Stahler (1993) predict that this occurs for stellar
masses in excess of 8~$M_\odot$. Once the star has ignited hydrogen burning,
further accretion should be inhibited by radiation pressure and powerful
stellar winds, with the consequence that stars more massive than 8~$M_\odot$
should not exist. Two formation scenarios have been proposed to solve this
paradox: non-spherical accretion (Yorke \& Sonnhalter 2002) and merging of
lower mass stars (Bonnell, Bate, \& Zinnecker 1998). Discriminating between these two
models represents a challenging observational goal.

In this context, the detection of disks would strongly favour the
accretion scenario, since random encounters between merging stars are not
expected to lead to axially symmetric structures. On the contrary,
conservation of angular momentum is bound to cause flattening and rotation of
the infalling material, thus producing disk-like bodies. Indeed,
circumstellar disks have been detected in low-mass stars and found to undergo
Keplerian rotation (Simon, Dutrey, \& Guilloteau 2000). Similar evidence has been found in a
few high-mass young stellar objects (YSOs), but in most cases the angular
resolution was insufficient to assess the presence of a disk unambiguously.
In conclusion, only few bona fide examples are known (see Cesaroni 2002) and
all of these are associated with moderately massive stars (B1-B0). This is
not sufficient to understand the role of disks in the formation of even more massive stars and establish the relevance of accretion to this
process.

With this in mind, we have decided to perform a search for disks in a limited
number of high-mass YSOs. For this purpose, we have selected two luminous
objects with typical signposts of massive star formation such as water masers
and ultracompact (UC) \HII\ regions. The first, G31.41+0.31 (hereafter G31), is
a well studied hot core located at 7.9~kpc (Olmi et al. 1996b; Cesaroni et al.
1998), where preliminary evidence of a rotating massive disk oriented
perpendicularly to a bipolar outflow has been reported in Cesaroni et al.
(1994). The other, G24.78+0.08 (hereafter G24), is a cluster of massive
(proto)stars with a distance of 7.7~kpc, where recently Furuya et al. (2002)
have detected a pair of cores, each of these associated with a compact bipolar
outflow. By analogy with G31.41 the expectation is that also in this case the
cores could contain rotating disks perpendicular to the flow axes.

On the basis of previous experience with this type of objects (see, e.g.,
Cesaroni et al. 1999), \MCN\ has been used as
disk tracer. This is a low-abundance molecule which is excited in
very dense regions. Therefore, searching for disks requires not only high
angular resolution, but also great sensitivity given the faintness of the
lines observed. In order to achieve these goals, we have used the Plateau
de Bure interferometer (PdBI) at 1.4~mm
in the most extended configuration. In this letter we present the discovery
of four rotating disks in the two regions studied and discuss possible
implications for the star formation process. A full report on the results
obtained and a more detailed analysis of the data will be presented in
a forthcoming article.

\section{Observations} 
\label{sobs}

We carried out observations in the 1.4~mm continuum and \MCN(12--11)
line emission with the PdBI on 2003 March 16. The
inner hole in the ($u,v$) plane has a radius of 15~$k\lambda$. Line data have been
smoothed to a spectral resolution of 0.5~\kms\ and channel maps were
created with natural weighting, attaining a
resolution of 1\farcs2$\times$0\farcs5 (full width at half power of
the synthesized beam) and a sensitivity of 40--50~mJy/beam/channel
(1$\sigma$ RMS). For the continuum map the resolution and sensitivity
are 1\farcs2$\times$0\farcs5 and 4--6~mJy/beam, respectively. 

In the following, we further analyze previous CS(3--2) observations
obtained with the Nobeyama Millimeter Array (NMA) by Cesaroni et
al. (2003), to whom we refer for technical details.

\section{Results and discussion}
\label{sres}

The main goal of our study was the discovery of rotating disks
associated with massive YSOs deeply embedded in dense, compact cores. This
was achieved searching for well defined {\it velocity gradients} in the
cores, {\it perpendicular to molecular outflows} powered by the YSOs, as
illustrated in the next sections.

\subsection{Structure of the cores}

The region G31 consists of a hot core, detected in various high-energy lines
(Olmi, Cesaroni, \& Walmsley 1996a) located at the center of a bipolar
outflow, at $\sim$5\arcsec\ from an UC \HII\ region (Cesaroni et al. 1998).
On the other hand, G24 is more complex, as it contains four distinct
objects (see Fig.~1 of Furuya et al. 2002): two of these, G24~A and G24~C,
are massive cores associated with two bipolar outflows and represent the
target of the present study.

\begin{figure}
\begin{center}
\includegraphics[angle=0,width=6.5cm]{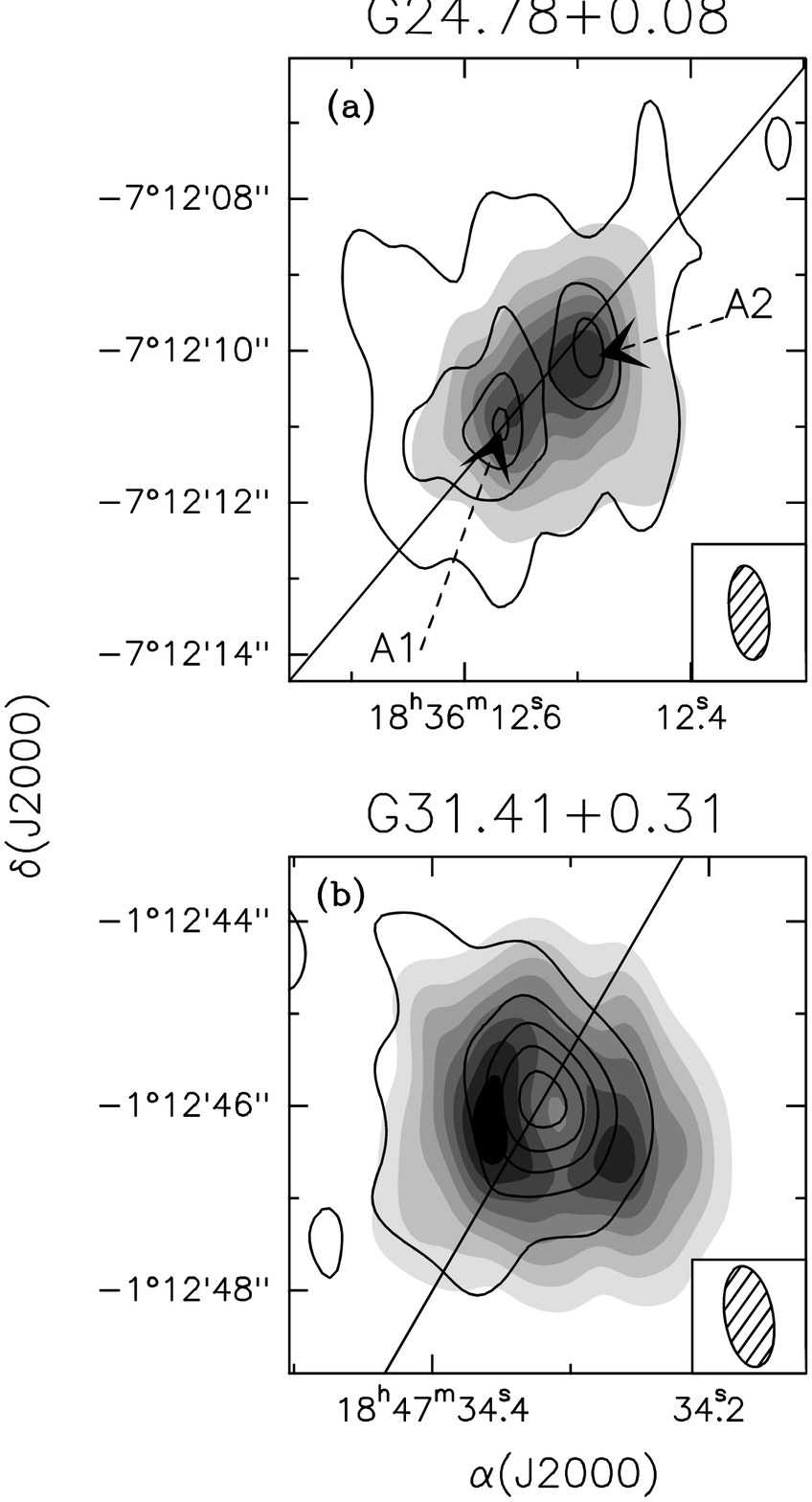}
\caption{{\bf Upper panel}: Overlay of the 1.4~mm continuum contour map
 on the gray scale image obtained integrating the \MCN(12--11)
 emission under the $K=0, 1$, and 2 components for G24. Contour levels range from
 0.02 to 0.2 in steps of 0.06~Jy\,beam$^{-1}$. Grayscale levels range from 0.1
 to 1.00 in steps of 
 0.18~Jy\,beam$^{-1}$\,\kms. The straight line represents the axis of the
 bipolar outflow. The synthesized beam is shown in the lower right-hand corner. {\bf Lower panel}: same as upper panel for G31. Contour levels
 range from 0.08 to 1.28 in steps of 0.3~Jy\,beam$^{-1}$. Grayscale levels range from 0.1 to 0.82 in 
 steps of 0.12~Jy\,beam$^{-1}$\,\kms.
}
\label{fcl}
\end{center}
\end{figure}

A picture of the G24~A and G31 cores is given
in Fig.~\ref{fcl}, where overlays of the 1.4~mm continuum and integrated
\MCN(12--11) line emission are shown.
Note that no map is shown for G24~C because no \MCN(12--11) line emission
has been detected with the PdBI. However, the 1.4~mm continuum flux is
consistent with the extrapolation of the spectral energy distribution
presented by Furuya et al. (2002), thus confirming the existence of such
core. Very likely the fact that G24~C is detected in the \MCN(8--7)
transitions (Furuya et al. 2002), but not in the (12--11) is due to this core
being significantly colder than G24~A (see Codella et al.\ 1997), which makes it difficult to detect
high energy lines.

When observed with sub-arcsec resolution, G24~A is resolved into two separate
cores. This is evident both in the 1.4~mm continuum and line maps. In the
following we shall refer to these cores as G24~A1 (the one to the SE) and
G24~A2 (the one to the NW). The former lies slightly closer to the geometrical
center of the bipolar outflow reported by Furuya et al. (2002), but the small
separation between the cores and the fact that they are aligned along the
outflow axis make it difficult to establish whether the outflow is indeed
associated with G24~A1; in the following we arbitrarily assume that this is the
case. The conclusions derived in our study are independent of  the association
of the outflow with either of the cores. Noticeably, G24~A1 coincides with an
unresolved UC \HII\ region detected by Codella, Testi, \& Cesaroni (1997),
whereas no free-free emission is reported towards G24~A2. To determine whether
this is an effect of different evolutionary stages of the two cores requires a
detailed comparison of their physical properties which we postpone to a
forthcoming paper.

The appearance of G31 is even more intriguing: while the 1.4~mm continuum seems to
trace a roughly spherical core, the \MCN\ map reveals a toroidal structure with
the dip centered at the position of the continuum peak (Fig.~\ref{fcl}). This suggests that
either the \MCN\ abundance drops dramatically in the central region of the
core, due to a temperature increase towards the center (in agreement with the
findings of Olmi et al. 1996b), or the temperature near to the embedded source
is so high that the ground level states of \MCN\ are under populated.
Interestingly, the two peaks of the \MCN\ emission are roughly symmetric with
respect to the axis of the outflow observed by Olmi et al. (1996b), suggesting
a physical connection between the toroid and the flow. Such a connection will
become more evident when considering the velocity field in the cores.

\subsection{Kinematics of the cores}

\begin{figure*}
\centerline{\includegraphics[angle=-90,width=16.0cm]{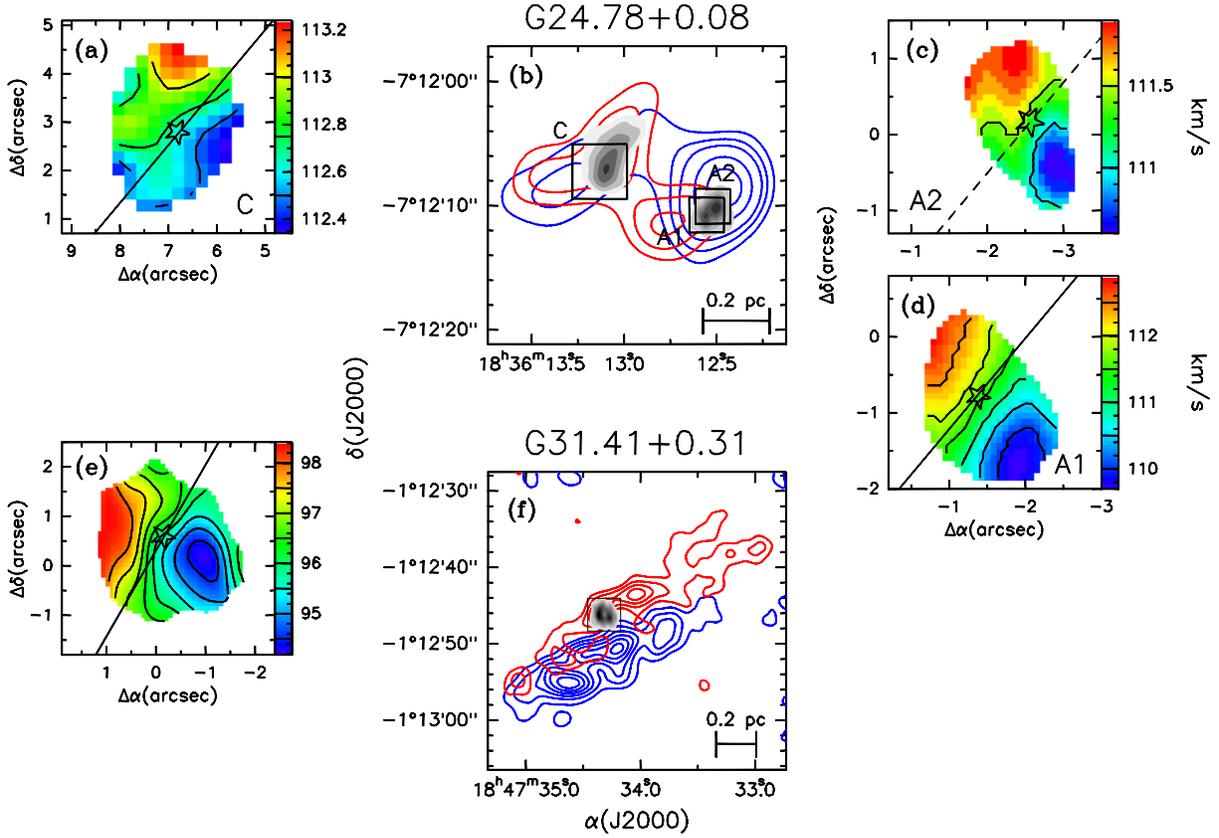}}
\caption{
 {\bf a.} Map of the first moment of the CS(3--2) line observed by
  Cesaroni et al. (2003) towards G24~C. The contour levels in \kms\ are indicated in the wedge to the right of the panel. The position offsets are relative to $\alpha(\mathrm{J2000})= 18^{\mathrm h} 36^{\mathrm m}
 12\fs66$, $\delta(\mathrm{J2000})= -07\degr 12\arcmin 10\farcs15$.
  The straight line represents the
  outflow axis. The star indicates the peak of the mm continuum emission.
 {\bf b.} Comparison between the bipolar outflows observed by
  Furuya et al. (2002) ({\it contours}), the \MCN(12--11) line emission mapped by us towards
  G24~A1 and G24~A2 ({\it gray scale}), and the CS(3--2) line map from Furuya et al. (2002)
  towards G24~C ({\it gray scale}). Grayscale levels for \MCN(12--11) are the same as in
  Fig.~\ref{fcl}, while for CS(3--2) grayscale levels range from 0.4 to 0.8 in steps of 0.1~Jy\,beam$^{-1}$\,\kms. Blue and red contours represent respectively the blue- and red-shifted $^{12}$CO(1--0) line emission (see Fig.~2 of Furuya et al.\ 2002).
 {\bf c.} Map of the \MCN(12--11) line peak velocity towards G24~A2 obtained with a Gaussian fit. The contour levels in \kms\ are indicated in the wedge to the right of the panel. The position offsets are relative to $\alpha(\mathrm{J2000})= 18^{\mathrm h} 36^{\mathrm m}
 12\fs66$, $\delta(\mathrm{J2000})= -07\degr 12\arcmin 10\farcs15$. The dashed line indicates the direction of the outflow, arbitrarily associated with G24~A1 (see text). The star indicates the peak of the mm continuum emission.
 {\bf d.} Same as c for G24~A1. 
 {\bf e.} Same as c for G31. The position offsets are relative to $\alpha(\mathrm{J2000})= 18^{\mathrm h} 47^{\mathrm m}
 34\fs33$, $\delta(\mathrm{J2000})= -01\degr 12\arcmin 46\farcs50$. 
 Note that for this case the direction of the outflow has been obtained connecting the peaks of the blue and red lobes.
 {\bf f.} Comparison between the bipolar outflow observed by Olmi et al. (1996b)
  ({\it contours}), and the \MCN(12--11) line emission map towards G31 ({\it gray scale}). Grayscale levels for \MCN(12--11) are the same as in
  Fig.~\ref{fcl}. Blue and red contours represent respectively the blue- and red-shifted $^{13}$CO(1--0) line emission (see Fig.~5 of Olmi et al.\ 1996b).
}
\label{fvel}
\end{figure*}

An obvious way to analyze the velocity field in the cores is to produce maps of
the line peak velocity obtained with Gaussian fits. Since multiple \MCN\
$K$-components are simultaneously observed in the same intermediate-frequency
bandwidth, it is possible to improve the accuracy of the fit by fitting all
lines together, assuming identical widths and fixing their separations to the
laboratory values (see, e.g., Olmi et al.\ 1993). Such a fit has been made in
each point where \MCN\ emission is detected. The maps of the LSR velocity for
the G24~A1, G24~A2, and G31 cores can be seen in Figs.~\ref{fvel}c,
\ref{fvel}d, and \ref{fvel}e. The same method could not be applied to G24~C
because the \MCN(12--11) line emission is not detected towards this core in our
PdBI observations, while the spectral resolution used for the \MCN(8--7)
transition by Furuya et al. (2002) was too poor (16~\kms). Hence, we have
re-analyzed the CS(3--2) data by Cesaroni et al. (2003), as the CS emission
line was much stronger and observed with sufficient spectral resolution
(0.5~\kms). In this case the line profile deviates significantly from a
Gaussian, presenting prominent emission in the red wing. Therefore we preferred
to estimate the velocity from the first moment\footnote{the first moment between $v_1$ and $v_2$ is defined as $\frac{\int^{v_2}_{v_1} T_B v dv}{\int^{v_2}_{v_1} T_B dv}$, where $T_B$ is the brightness temperature.} computed
over a velocity interval including only the peak of the emission, from 108 to
116~\kms, rather than from a Gaussian fit.
The resulting \Vlsr\ map is shown in Fig.~\ref{fvel}a.

The first conclusion that can be drawn from this figure is that all cores show
clear velocity gradients, with \Vlsr\ increasing steadily along well defined
directions. We examine three possible explanations for such gradients:
expansion, infall, or rotation. The first can be ruled out as the velocity
gradient should be maximum in the same direction as the molecular outflow,
which is clearly not the case (see Figs.~\ref{fvel}b and \ref{fvel}e).
Spherical infall is also impossible, because self-absorption would shift the
peak velocity towards lower values at the core center, whereas we observe a
steady velocity increase along a well defined direction (see Figs.~\ref{fvel}a,
\ref{fvel}c, \ref{fvel}d, and \ref{fvel}e). The fact that such a direction is
perpendicular to the outflow axis strongly favours the rotation hypothesis:
this is exactly what one expects if the core is rotating about the axis of the
corresponding outflow. This behaviour might be mimicked also by two distinct
cores with different \Vlsr\ and too close to be resolved by our observations:
the velocity gradient would be a consequence of line emission from the two
cores observed in the same instrumental beam. However, we believe this to be
very unlikely. In fact, at least in the case of G31, the angular separation
between the regions emitting at the maximum and minimum velocities is
definitely greater than the beam size. As for G24, the two cores A1 and A2 likely correspond to two distinct rotating disks. However, one cannot rule out another explanation, namely that they are part of the same geometrically thick rotating disk: in this case, the two emission peaks would be produced by the interaction of the outflow with the dense material of the disk.

In conclusion, we believe that the most plausible explanation for
the kinematics of G24 and G31 is that {\bf the cores have toroidal structures
undergoing rotation about the corresponding outflow axis}. Hereafter, we shall
refer to these simply as {\bf disks}, although one has to keep in mind that
these are very different from the geometrically thin circumstellar disks seen
in low-mass YSOs.

\subsection{Nature of the G24 and G31 disks}

The major question raised by our results is whether the disks are stable
entities. In Table~\ref{tpar} we give a few disk parameters, among which the
mass of the cores, $\Mgas$, and the dynamical mass, $\Mdyn$,
needed for equilibrium. The former was estimated from the millimeter continuum emission assuming a mass
opacity of $\simeq 0.02$~cm$^{-2}$\,g$^{-1}$ at 1.4~mm for a gas-to-dust ratio
of 100 (see, e.g., Andr\'e, Ward-Thompson, \& Barsony 2000), and the
temperatures listed in Table~\ref{tpar}; the latter was computed assuming equilibrium between centrifugal and
gravitational forces from the expression $\Mdyn=\mbox{$v^2_{\rm rot}$}\,R\sin^2i/G$, where \vrot\ is the rotation velocity, $R$ is the radius of the disk, and $i$
is the inclination angle of the disk assumed to be $45\degr$. $\Mgas$
is much larger than $\Mdyn$, suggesting that the disks may be unstable. In
principle, magnetic fields could stabilize the disks, but this would require a
few 20--40~mG, values too large to be plausible even in regions as dense as
$10^8$~cm$^{-3}$ (see Fig.~1 of Crutcher 1999). Therefore, the disks must be
transient structures with lifetimes of the order of the free-fall time, \tff,
also listed in Table~\ref{tpar}. Another estimate of the disks lifetime, \tacc,
can be derived from the ratio between the disk mass, \Mgas, and the accretion
rate. The latter is computed from the expression $\Macc=2\pi\Sigma R \vin$
where $\Sigma=\Mgas/\pi R^2$ is the surface density and \vin\ is the infall
velocity, which has been assumed to be equal to the rotation velocity following
Allen, Li, \& Shu (2003). As one can see from Table~\ref{tpar}, \tff\ is very close
to \tacc, and both agree within a factor $\le$4 with the outflow age, \tout,
derived from the data of Olmi et al. (1996b) and Furuya et al. (2002). Note
that \tout\ is to be multiplied by $\cot\theta$ to correct for the (unknown)
inclination angle $\theta$ of the flow with respect to the line of sight. We
believe that $\theta$ cannot differ significantly from $45\degr$ otherwise
blue- and red-shifted emission would mix up in the plane of the sky (for
$\theta\simeq90\degr$) or along the line of sight through the center (for
$\theta\simeq0\degr$): this implies a correction factor of order unity. The
correction factors for $\theta=30\degr$ and $60\degr$ would be 1.7 and 0.6
respectively.

In conclusion, the lifetime of the disks seems to be of order of $10^4$~yr.
Such a short lifetime should imply 10 times less disks than UC \HII\ regions,
which are supposed to live $10^5$~yr (Wood \& Churchwell 1989). Although it is obviously impossible to
confirm this estimate on a statistical ground, disks appear to be an ubiquitous
phenomenon in massive star forming regions, as we have detected 4 of them in
2 regions only. Therefore, it seems unlikely that disks are 10 times
less numerous than UC \HII\ regions. This implies a significantly longer
lifetime than $\sim$$10^4$~yr, which in turn means that disks must be fed by
a larger scale reservoir of material at a rate comparable to
$\Macc\simeq10^{-2}~M_\odot$~yr$^{-1}$. Accretion rates that large have been
estimated by Fontani et al. (2002) for the parsec-scale clumps where
high-mass star formation is observed.

If the disk lifetime is comparable to that of UC \HII\ regions, then
the total accreted mass should result in
$\Macc\times10\,\tff\simeq10^{-2}\times10^5=10^3~M_\odot$ of stars, too large
a value for a single star, but acceptable if the infalling gas is accreting
onto a cluster of stars. Indeed, this resembles the situation in the Orion cluster (Palla
\& Stahler 1999), where only $<$7\% of the mass in stars ($>$600~$M_\odot$) is
contained in the most massive star of the cluster ($\sim$40~$M_\odot$).

The main finding of our study is that we have detected rotating disks
associated with high-mass YSOs, and  hence, this result strongly suggests that
non-spherical accretion is a viable mechanism to form high-mass stars. Only a
larger number of observations may confirm this conclusion on a statistical
ground, thus assessing that disks are a natural product of the star formation
process also for early-type stars.

\acknowledgements
It is a pleasure to thank the staff of IRAM
for his help during the observations. We also thank Daniele
Galli for stimulating discussions about models of massive accretion disks.


\clearpage

\clearpage

\clearpage

\begin{deluxetable}{lcccccccccc}
\tabletypesize{\scriptsize}
\tablecaption{Parameters of disks and outflows in G24 and G31.
\label{tpar}}
\tablehead{
\colhead{Core} & \colhead{$T$} &\colhead{$R$} &\colhead{\vrot} &\colhead{\Mdyn} &\colhead{\Mgas} &\colhead{\tff} &\colhead{\tacc} &\colhead{\tout} &\colhead{\Macc\tablenotemark{\phd{\rm a}}} &\colhead{\Mout} 
\\
&\colhead{(K)} &\colhead{(pc)} &\colhead{(\kms)} &\colhead{($M_\odot$)} &\colhead{($M_\odot$)} &\colhead{(yr)} &\colhead{(yr)} &\colhead{(yr)} &\colhead{($M_\odot$yr$^{-1}$)} &\colhead{($M_\odot$yr$^{-1}$)} 
}
\startdata
G24 A1 & ~~80\tablenotemark{b} & 0.02 & 1.50 &  23 & 130 & $4\times10^3$ & $7\times10^3$ & $2\times10^4$\tablenotemark{\phd{\rm c}} & $2\times10^{-2}$ & $5\times10^{-4}$\tablenotemark{\phd{\rm c}} \\
G24 A2 & ~~80\tablenotemark{b} & 0.02 & 0.75 &   4 & ~80 & $3\times10^3$ & $1\times10^4$ &   ---        & $8\times10^{-3}$ &    ---           \\
G24 C  & ~~30\tablenotemark{b} & 0.04 & 0.50 &   5 & 250 & $9\times10^3$ & $4\times10^4$ & $2\times10^4$\tablenotemark{\phd{\rm c}} & $6\times10^{-3}$ & $5\times10^{-4}$\tablenotemark{\phd{\rm c}} \\
G31    & ~230\tablenotemark{d} & 0.04 & 2.10 &  87 & 490 & $6\times10^3$ & $1\times10^4$ & $2\times10^5$\tablenotemark{\phd{\rm d}} & $5\times10^{-2}$ & $4\times10^{-4}$\tablenotemark{\phd{\rm d}} \\
\enddata
\tablenotetext{a}{Computed assuming \vin=\vrot.}
\tablenotetext{b}{ From Codella et al. (1997).} 
\tablenotetext{c}{ From Furuya et al. (2002).}
\tablenotetext{d}{ From Olmi et al. (1996b).} 
\end{deluxetable}

\end{document}